# Anomalous Lattice Vibrations of Single and Few-Layer MoS$_2$

**Changgu Lee[1], Hugen Yan[2], Louis E. Brus[3], Tony F. Heinz[2,4], James Hone[1], Sunmin Ryu[5]***

[1]Department of Mechanical Engineering, Columbia University, New York, NY 10027
[2]Department of Physics, Columbia University, New York, NY 10027
[3]Department of Chemistry, Columbia University, New York, NY 10027
[4]Department of Electrical Engineering, Columbia University, New York, NY 10027
[5]Department of Applied Chemistry, Kyung Hee University, Yongin, Gyeonggi 446-701, Korea
*E-mail: sunryu@khu.ac.kr

**Abstract**

Molybdenum disulfide (MoS$_2$) of single and few-layer thickness was exfoliated on SiO$_2$/Si substrate and characterized by Raman spectroscopy. The number of S-Mo-S layers of the samples was independently determined by contact-mode atomic-force microscopy. Two Raman modes, E$^1_{2g}$ and A$_{1g}$, exhibited sensitive thickness dependence, with the frequency of the former decreasing and that of the latter increasing with thickness. The results provide a convenient and reliable means for determining layer thickness with atomic-level precision. The opposite direction of the frequency shifts, which cannot be explained solely by van der Waals interlayer coupling, is attributed to Coulombic interactions and possible stacking-induced changes of the intralayer bonding. This work exemplifies the evolution of structural parameters in layered materials in changing from the 3-dimensional to the 2-dimensional regime.

**Main text**

Recent advances in the formation of atomically thin layers of van der Waals bonded solids by mechanical exfoliation have opened up important new possibilities for the investigation of two-dimensional systems.[1] Single-layer graphene, for example, exhibits many distinctive physical properties not found in its bulk counterpart,graphite.[2] Further, the ability to form samples with a controlled number of atomic layers permits one to examine precisely the evolution of material properties with thickness. Because the vibrational spectrum is sensitive to the sample thickness, Raman spectroscopy has been widely used to determine the number of layers, as well as to examine the changes in material properties with thickness.[3-5] A largely unexplored question, however, is how weak van der Waals (vdW)-like interlayer interactions affect the intralayer bonding and lattice vibrations of stacked few-layer samples. In bulk graphite, this weak interaction leads to rigid-layer vibrations at very low frequency ($E_{2g}$ at 42 cm$^{-1}$) and a negligible splitting between conjugate modes ($E_{2g}$ and $E_{1u}$ both at 1582 cm$^{-1}$).[6] Experimental investigations of interlayer interactions in few-layer graphene samples have, however, been hampered by the sensitivity of the vibrations to environmental charge doping through the presence of strong non-adiabatic electron-phonon interactions.[7-9]

Molybdenum disulfide ($MoS_2$), a naturally occurring as molybdenite, is one of the most stable layered metal dichalcogenides.[10, 11] It is built up from layers consisting of an atomic plane of Mo sandwiched between two atomic planes of S in a trigonal prismatic arrangement.[12] In the bulk, it is a semiconductor with a direct (indirect) band gap of 1.96 eV (1.2 eV)[13] and therefore does not exhibit the complexities associated with the non-adiabatic electron-phonon coupling present in graphene. The material has attracted interest for diverse applications. In addition to its role as a catalyst in dehydrosulfurization,[10] the layered structure of $MoS_2$ allows its use as a dry lubricant.[11] The material has also been investigated for photovoltaic power generation and photo-electrochemical hydrogen production.[14-16] A recent study demonstrated that nanoclusters of $MoS_2$ can significantly enhance photocatalytic hydrogen production when it is mixed with CdS.[17] Because of its capability to form intercalation compounds,[18] $MoS_2$ is an ingredient of some Li ion batteries.[19] As exemplified by the recent studies on few-layer graphene films, its physical properties should depend sensitively on the number of layers in atomically thin samples of $MoS_2$.[20] However, no systematic studies of these effects have yet been reported for atomically thin films, although researchers have identified interesting size-dependent properties in $MoS_2$ nanoparticles (20 - 200 nm in diameter)[21] and nanoribbons (3 - 100 nm in thickness).[22]

In this article, we present the first report of Raman scattering studies of single and few-layer $MoS_2$ samples. We observe clear signals of the in-plane ($E^1_{2g}$) and out-of-plane Raman modes for samples of 1, 2, 3, …, 6 layers. We find that these modes exhibit a well-defined thickness dependence, with the two modes shifting away from each other in frequency with increasing thickness. This provides a convenient and precise diagnostic for determination of the layer thickness of samples. The behavior of the frequency shifts with layer thickness cannot be explained solely in terms of weak vdW interlayer interactions. We also observed a distinctive thickness dependence of the intensities and linewidths of both Raman modes.

Single and few-layer $MoS_2$ films were deposited from bulk $MoS_2$ crystals onto Si substrates covered with a 285-nm-thick $SiO_2$ layer. The $MoS_2$ layers were formed by the mechanical exfoliation method widely employed for the deposition of graphene samples.[1] Atomically thin films were first identified by optical microscopy. They were then imaged by atomic-force microscopy (AFM) to determine the sample thickness (Fig.1). The contact mode of measurement (in the repulsive force regime) was chosen, since it was recently shown to produce better accuracy in the characterization of graphene samples than the noncontact or tapping-mode of operation.[23] For $MoS_2$ samples exhibiting lateral thickness variation, we observed step heights of individual layers of 0.6 - 0.7 nm. This value is compatible with the 0.62-nm interlayer spacing of a single layer of the S-Mo-S building block of the MoS$_2$ crystal.[24] The measurements show that exfoliation produces layers with a discrete number of these units. We consequently designate the thickness of our films in terms of the number of these MoS$_2$ layers (nL). In the AFM measurements, single-layer films on the bare substrates showed a wider distribution in heights (0.6 - 0.9 nm). This may reflect the presence of adsorbates below the film or other interactions between the film and oxide substrate surface.[23] From extensive AFM scanning of freshly deposited samples, we found no evidence of structural irregularity on the nanometer length scale.

Figure 2 shows representative Raman spectra for single and few-layer MoS$_2$ samples. Among the four Raman-active modes of the bulk 2H-MoS$_2$ crystal (Fig. 2(e)),[12] we observed the $E^1_{2g}$ and $A_{1g}$ modes near 400 cm$^{-1}$. The other two modes ($E_{1g}$, $E^2_{2g}$) could not be detected either because of selection rules for our scattering geometry ($E_{1g}$)[12] or because of the limited rejection of the Rayleigh scattered radiation ($E^2_{2g}$).[25] We find that single-layer MoS$_2$ exhibits a strong in-plane vibrational mode at ~384 cm$^{-1}$, corresponding to the $E^1_{2g}$ mode of the bulk 2H-MoS$_2$ crystal. In contrast, this mode was not observed in earlier studies of single-layer MoS$_2$ suspensions prepared from Li-intercalated MoS$_2$.[26] Its absence was attributed to the existence of metastable octahedral coordination, which renders the vibration Raman inactive ($E_u$).[26] Our data indicate that exfoliated single layers maintain the trigonal prismatic coordination of bulk MoS$_2$ and also suggest that the octahedral coordination found in dispersed single layers arises from an intercalation-assisted phase transformation.[26]

For all film thicknesses, the Raman spectra in Fig. 2 show strong signals from both the in-plane $E^1_{2g}$ and the out-of-plane $A_{1g}$ vibration. The behavior as function of film thickness has several intriguing characteristics. Most strikingly, we find (Fig. 2(a-b)) that the $E^1_{2g}$ vibration softens (redshifts), while the $A_{1g}$ vibration stiffens (blueshifts) with increasing sample thickness. For films of four or more layers, the frequencies of both modes converge to the bulk values. The rate of frequency change is twice as large as for the $A_{1g}$ as for $E^1_{2g}$ mode. Spatial maps of the Raman frequency for the $E^1_{2g}$ (Fig. 2(c)) and $A_{1g}$ mode (Fig. 2(d)) clearly show these opposing shifts with layer thickness. The maps also demonstrate that the frequencies of the two modes have only very slight variation for different locations within a sample of a given layer thickness. This permits the Raman frequencies to be used as an indicator of the layer thickness. The opposite variation with layer thickness of the frequencies of these two Raman modes renders the difference in their frequencies (Δω) a particularly effective 'thickness indicator.' For single MoS$_2$ layers, for example, Δω is ~3 cm$^{-1}$ smaller than the value of Δω for bilayers. This change is several times larger than the typical variation of Δω within or between different single-layer samples (Fig. 2(b)). The intensities and linewidths of the two modes also show distinctive variation as a function of film thickness, as we discuss below.

The vibrations of bulk materials built up from van der Waals bonded layers are often analyzed in terms of the two-dimensional layers from which they are formed.[12, 27] Many approaches have been developed within this weak coupling limit to describe the relation between the vibrational modes within thin layers and the those of bulk material.[12, 24, 27, 28] Within a classical model for coupled harmonic oscillators,[20] the $E^1_{2g}$ and $A_{1g}$ modes are expected to stiffen as additional layers are added to form the bulk material from individual layers, since the interlayer vdW interactions increase the effective restoring forces acting on the atoms. While the shift of $A_{1g}$ mode observed in our measurements with increasing layer number agrees with this prediction, the behavior of the $E^1_{2g}$ mode does not. The failure of the model could reflect the presence of additional interlayer interactions; it could also indicate that the implicit assumption that stacking does not affect intralayer bonding is incorrect. Regarding the latter, a low-energy electron diffraction study of $MoS_2$ single crystals[29] showed that the interplane distance between Mo and S atomic planes within the topmost layer shrinks by ~5% compared to its bulk value. The lateral lattice expansion observed for dispersed single layers[26] may also be related to surface reconstruction. In addition, the $A_{1g}$ mode of the topmost layer of bulk $MoS_2$ crystals[30] was found to soften by 25 cm$^{-1}$, possibly because of the surface reconstruction. The observed surface reconstruction and vibrational softening show that even the nominally weak interlayer interaction in $MoS_2$ can affect intralayer bonding and lattice dynamics. The same considerations may apply to single and few-layer $MoS_2$ samples.

The shift in the frequency of the $A_{1g}$ mode as a function of thickness in the current study is consistent with the transition from surface to bulk layers.[30] The opposite progression for the $E^1_{2g}$ mode may reflect the influence of stacking-induced structural changes. Alternatively, the anomalous behavior of $E^1_{2g}$ may be attributed to long-range Coulombic interlayer interactions.[28] Indeed, a similar anomaly was observed for the $E^1_{2g}$ and $E_{1u}$ conjugate in-plane modes of bulk $MoS_2$,[12] which would be degenerate in the absence of interlayer interactions. According to the weak coupling model, the symmetric $E^1_{2g}$ should have a higher frequency than the antisymmetric (IR active) $E_{1u}$ mode, since the latter is barely affected by interlayer interactions.[31] Experimentally, however, the opposite was observed.[31] Such an anomalous Davydov splitting[32] has also been observed in GaSe[28] and GaS.[33] A non-negligible ionic interlayer interaction between the metallic and chalcogen atoms has been invoked to explain these results.[28, 31, 32]

Another seemingly perplexing result is that the Raman intensities from single layers are stronger than from bulk $MoS_2$ samples (Fig 3(a)). With increasing thickness, the intensities rise roughly linearly up to four layers and then decrease for thicker samples. A similar enhancement in the Raman intensity has been reported for graphene deposited on $SiO_2$/Si substrates, where the maximum intensity occurs for ~10 layers.[34] As in the case of the graphene layers,[34] much of this effect is clearly attributable simply to optical interference occurring for both the excitation laser and the emitted Raman radiation. The presence of the oxide film above the silicon substrate leads to significant optical field enhancements. The importance of this optical interference effect is highlighted by fact that single-layer $MoS_2$ on quartz produced only 20% of the Raman signal observed for such samples deposited on the $SiO_2$/Si substrates. In addition, changes in the electronic structure of the $MoS_2$ samples with layer thickness[35] may also play a role in the observed variation of Raman intensities.

Fig. 3(b) presents linewidths ($\Gamma$) of both of the Raman modes as a function of layer thickness. While $\Gamma[E^1_{2g}] \approx 2.1$ cm$^{-1}$ is largely independent of thickness, we find that $\Gamma[A_{1g}]$ increases from 3 to 6 cm$^{-1}$ as the thickness decreases from 6 to 2 layers. The increase of $\Gamma[A_{1g}]$ with decreasing thickness may reflect the presence of varying force constants associated with structural changes between the inner and outer layers of the material. For thicker layers, the bulk-like inner layers will begin to dominate the Raman intensity, leading to a reduced linewidth. Among other factors that might contribute to the observed broadening, we can exclude lateral sample inhomogeneity (including in-plane strain). We find that $\Gamma[A_{1g}]$ is larger for bilayers than for single layers, while single layers show the greatest frequency variation among different samples (Fig. 2b). We are also able to exclude any significant effect of interactions of the substrate with our atomically thin $MoS_2$ samples. To understand the influence of the underlying $SiO_2$/Si substrates on the observed phonon dynamics, freestanding $MoS_2$ layers (1L-4L in thickness) were studied in comparison with $MoS_2$ layers supported on $SiO_2$. (See Methods for details.) We examined freestanding $MoS_2$ films suspended over micron-sized holes (Fig. 4a). As shown in Fig. 4b, we did not observe meaningful differences in the Raman frequencies and linewidths of either mode compared with the samples on the oxide surface. (See Supporting Information for a comparison of the linewidths.)

The reduced sensitivity of $\Gamma[E^1_{2g}]$ to thickness can be attributed to the less efficient interlayer coupling of the in-plane $E^1_{2g}$ mode.[36] The fact that the maximum of $\Gamma[A_{1g}]$ occurs for bilayers rather than for single layers may reflect the intrinsic differences in symmetry between single and bilayer samples.[37] The ratio of the integrated Raman intensities for the two modes (Fig. 3(a)) also shows distinctive behavior for single-layer samples. The origin of this difference is unclear, but it should not arise from the optical effects described above, since they will affect both modes almost identically. It also does not appear to arise from the existence of phases lacking trigonal prismatic coordination, such as the octahedrally coordinated metastable phase. The presence of such phases would reduce the $E^1_{2g}$ Raman response. However, we found no change in the signal when annealing the samples to 100 - 300 $^o$C, a procedure which should accelerate any possible phase transformation.[37] Differences in the electronic structure of the samples as a function of layer thickness[35] may, however, also contribute to the change in the ratio of the Raman response.

In conclusion, we have characterized single and few-layer $MoS_2$ films by AFM and Raman spectroscopy. We observed that the frequency of the $E^1_{2g}$ mode decreases and that of $A_{1g}$ mode increases with increasing layer thickness. Since the sign of the shift of the $E^1_{2g}$ mode is unexpected within models of weak vdW interlayer coupling, the results suggest a role for stacking-induced changes in intralayer bonding and/or the presence of Coulombic interlayer interactions in $MoS_2$. We show that the frequency difference between $E^1_{2g}$ and $A_{1g}$ modes can be used as a robust and convenient diagnostic of the layer thickness of $MoS_2$ samples. The present study should assist future investigations of the electronic, optical, chemical, and mechanical properties of atomically thin $MoS_2$ films.

**Methods**

***Preparation of single and few-layer $MoS_2$ films***

Single and few-layer $MoS_2$ films were isolated from bulk crystals of 2H-$MoS_2$ (SPI, natural molybdenite) using the micro-mechanical exfoliation method widely adopted for preparation of graphene samples.[1] In a typical run, a small piece of bulk $MoS_2$ was put on the sticky side of a piece of adhesive tape. This was followed by repeated folding and unfolding of the tape to produce thin flakes along the tape's surface. The tape was then pressed onto a silicon substrate covered by a 285 nm-thick $SiO_2$ layer. After mild rubbing, we removed the tape from the substrate, leaving behind some atomically thin $MoS_2$ flakes. The size of single layers ranged from 1 to 10 μm, dimensions somewhat smaller than are typical for single-layer graphene exfoliated from kish graphite. The overall yield of atomically thin films of $MoS_2$ was, however, comparable to that for the production of graphene. As can be seen in Fig. 1a, even $MoS_2$ films of single-layer thickness provided high optical visibility on $SiO_2$/Si substrates. The contrast decreased somewhat for the substrates used for the suspended samples, which had slightly thicker (~300 nm) epilayers of $SiO_2$.

*Preparation of suspended samples*

The substrates for suspended samples were fabricated using nanoimprint lithography followed by reactive-ion etching (RIE). Using electron-beam lithography, a 5 × 5 mm array of circles of 1 and 1.5 μm in diameter was patterned onto a nanoimprint master. The circles were then etched to a depth of 100 nm using RIE. A chromium layer of 10 nm thickness was deposited onto a Si wafer with a 300-nm $SiO_2$ epilayer by thermal evaporation. Before imprinting arrays of micron-sized circular holes, the Si wafer was coated by 120-nm thick PMMA layer. The imprint master coated with an anti-adhesion agent (NXT-110, Nanonex) was then used to imprint the PMMA to a depth of 100 nm. The thin PMMA layers remaining in the imprinted circular areas were etched by oxygen RIE; the exposed chromium layer was removed using a chromium etchant (Cyantek Cr-7S). To etch through the oxide and into the silicon to a total depth of 500nm, fluorine-based RIE was exploited using the remaining chromium layer as an etch mask. Following the last RIE step, the chromium etch mask was removed.

*Raman spectroscopy*

Analysis of the samples by micro-Raman spectroscopy was performed under ambient conditions. The pump radiation was supplied by an Ar-ion laser operating at a wavelength of 514.5 nm; the Raman emission was collected by a 40x objective in a backscattering geometry. The instrumental spectral resolution was 2.7 $cm^{-1}$, and the Si Raman band at 520 $cm^{-1}$ was used as an internal frequency reference. The Raman measurements were reproducible, as checked by characterization of ~30 single and few-layer samples. The Raman characterization of suspended samples was carried out with a different micro-Raman instrument, which was operated at an excitation wavelength of 532 nm and a spectral resolution of 1.0 $cm^{-1}$.

**Acknowledgements**

This research was supported by Basic Science Research Program through the National Research Foundation of Korea (NRF) funded by the Ministry of Education, Science and Technology (2009-0066575). Support for Columbia participants was provided by the US Department of Energy EFRC program (grant DE-SC00001085) and the Office of Basic Energy Sciences (grants DE FG 02-98ER14861 and DE FG 03ER15463) and by the New York State NYSTAR program. The authors appreciate helpful discussions with Prof. Aron Pinczuk.

**Supporting Information Available**

Detailed linewidth analysis of Raman spectra of freestanding $MoS_2$ films compared to films supported on $SiO_2$/Si substrates. This material is available free of charge *via* the Internet at http://pubs.acs.org.

**References**

1. Novoselov, K. S.; Jiang, D.; Schedin, F.; Booth, T. J.; Khotkevich, V. V.; Morozov, S. V.; Geim, A. K., Two-Dimensional Atomic Crystals. *Proc. Natl. Acad. Sci. USA* **2005,** *102*, 10451-10453.
2. Geim, A. K.; Novoselov, K. S., The Rise of Graphene. *Nat. Mater.* **2007,** *6*, 183-191.
3. Ferrari, A. C.; Meyer, J. C.; Scardaci, V.; Casiraghi, C.; Lazzeri, M.; Mauri, F.; Piscanec, S.; Jiang, D.; Novoselov, K. S.; Roth, S.; Geim, A. K., Raman Spectrum of Graphene and Graphene Layers. *Phys. Rev. Lett.* **2006,** *97*, 187401/1-187401/4.
4. Gupta, A.; Chen, G.; Joshi, P.; Tadigadapa, S.; Eklund, P. C., Raman Scattering from High-Frequency Phonons in Supported N-Graphene Layer Films. *Nano Lett.* **2006,** *6*, 2667-2673.
5. Graf, D.; Molitor, F.; Ensslin, K.; Stampfer, C.; Jungen, A.; Hierold, C.; Wirtz, L., Spatially Resolved Raman Spectroscopy of Single- and Few-Layer Graphene. *Nano Lett.* **2007,** *7*, 238-242.
6. Reich, S.; Thomsen, C., Raman Spectroscopy of Graphite. *Phil. Trans. R. Soc. Lond. A* **2004,** *362*, 2271–2288.
7. Yan, J.; Zhang, Y.; Kim, P.; Pinczuk, A., Electric Field Effect Tuning of Electron-Phonon Coupling in Graphene. *Phys. Rev. Lett.* **2007,** *98*, 166802/1-166802/4.
8. Pisana, S.; Lazzeri, M.; Casiraghi, C.; Novoselov, K. S.; Geim, A. K.; Ferrari, A. C.; Mauri, F., Breakdown of the Adiabatic Born-Oppenheimer Approximation in Graphene. *Nat. Mater.* **2007,** *6*, 198-201.
9. Casiraghi, C.; Pisana, S.; Novoselov, K. S.; Geim, A. K.; Ferrari, A. C., Raman Fingerprint of Charged Impurities in Graphene. *Appl. Phys. Lett.* **2007,** *91*, 233108/1-233108/3.
10. Heising, J.; Kanatzidis, M. G., Exfoliated and Restacked $MoS_2$ and $WS_2$: Ionic or Neutral Species? Encapsulation and Ordering of Hard Electropositive Cations. *J. Am. Chem. Soc.* **1999,** *121*, 11720-11732.
11. Kim, Y.; Huang, J.-L.; Lieber, C. M., Characterization of Nanometer Scale Wear and Oxidation of Transition Metal Dichalcogenide Lubricants by Atomic Force Microscopy. *Appl. Phys. Lett.* **1991,** *59*, 3404-3406.
12. Verble, J. L.; Wieting, T. J., Lattice Mode Degeneracy in $MoS_2$ and Other Layer Compounds. *Phys. Rev. Lett.* **1970,** *25*, 362-365.

13. Frey, G. L.; Elani, S.; Homyonfer, M.; Feldman, Y.; Tenne, R., Optical-Absorption Spectra of Inorganic Fullerenelike $MS_2$ (M=Mo, W). *Phys. Rev. B* **1998,** *57*, 6666-6671.
14. Evans, E. L.; Griffiths, R. J. M.; Thomas, J. M., Kinetics of Single-Layer Graphite Oxidation: Evaluation by Electron Microscopy. *Science* **1971,** *171*, 174.
15. Fortin, E.; Sears, W. M., Photovoltaic Effect and Optical Absorption in Mos2. *Journal of Physics and Chemistry of Solids* **1982,** *43*, 881-884.
16. Scrosati, B., Semiconductor Materials for Liquid Electrolyte Solar Cells. *Pure and Applied Chemistry* **1987,** *59*, 1173-1176.
17. Zong, X.; Yan, H.; Wu, G.; Ma, G.; Wen, F.; Wang, L.; Li, C., Enhancement of Photocatalytic $H_2$ Evolution on CdS by Loading $MoS_2$ as Cocatalyst under Visible Light Irradiation. *J. Am. Chem. Soc.* **2008,** *130*, 7176-7177.
18. Divigalpitiya, W. M. R.; Frindt, R. F.; Morrison, S. R., Inclusion Systems of Organic Molecules in Restacked Single-Layer Molybdenum Disulfide. *Science* **1989,** *246*, 369-371.
19. Schalkwijk, W. A. v.; Scrosati, B., *Advances in Lithium-Ion Batteries*. Kluwer Academic Publishers: Norwell, 2002.
20. Li, T.; Galli, G., Electronic Properties of $MoS_2$ Nanoparticles. *J. Phys. Chem. C* **2007,** *111*, 16192-16196.
21. Frey, G. L.; Tenne, R.; Matthews, M. J.; Dresselhaus, M. S.; Dresselhaus, G., Raman and Resonance Raman Investigation of $MoS_2$ Nanoparticles. *Phys. Rev. B* **1999,** *60*, 2883-2892.
22. Li, Q.; Walter, E. C.; Veer, W. E. v. d.; Murray, B. J.; Newberg, J. T.; Bohannan, E. W.; Switzer, J. A.; Hemminger, J. C.; Penner, R. M., Molybdenum Disulfide Nanowires and Nanoribbons by Electrochemical/Chemical Synthesis. *J. Phys. Chem. B* **2005,** *109*, 3169-3182.
23. Nemes-Incze, P.; Osva´th, Z.; Kamara´s, K.; Biro, L. P., Anomalies in Thickness Measurements of Graphene and Few Layer Graphite Crystals by Tapping Mode Atomic Force Microscopy. *Carbon* **2008,** *46*, 1435-1442.
24. Wakabayashi, N.; Smith, H. G.; Nicklow, R. M., Lattice Dynamics of Hexagonal $MoS_2$ Studied by Neutron Scattering. *Phys. Rev. B* **1975,** *12*, 659-663.
25. Verble, J. L.; Wieting, T. J.; Reed, P. R., Rigid-Layer Lattice Vibrations and van der Waals Bondings in Hexagonal $MoS_2$. *Solid State Commun.* **1972,** *11*, 941-944.
26. Yang, D.; Sandoval, S. J.; Divigalpitiya, W. M. R.; Irwin, J. C.; Frindt, R. F., Structure of Single Molecular Layer $MoS_2$. *Phys. Rev. B* **1991,** *43*, 12053-12056.
27. Bromley, R., Lattice Vibrations of $MoS_2$ Structure. *Philos. Mag.* **1971,** *23*, 1417-1427.
28. Wieting, T. J.; Verble, J. L., Interlayer Bonding and the Lattice Vibrations of Beta-GaSe. *Phys. Rev. B* **1972,** *5*, 1473-1479.
29. Mrstik, B. J.; Kaplan, R.; Reinecke, T. L.; Hove, M. V.; Tong, S. Y., Surface-Structure Determination of the Layered Compounds $MoS_2$ and $NbSe_2$ by Low-Energy Electron Diffraction. *Phys. Rev. B* **1977,** *15*, 897-900.
30. Bertrand, P., Surface-Phonon Dispersion of $MoS_2$. *Phys. Rev. B* **1991,** *44*, 5745-5749.
31. Ghosh, P. N.; Maiti, C. R., Interlayer Force and Davydov Splitting in 2H-$MoS_2$. *Phys. Rev. B* **1983,** *28*, 2237-2239.
32. Ghosh, P. N., Davydov Splitting and Multipole Interactions. *Solid State Commun.* **1976,** *19*, 639-642.
33. Kuroda, N.; Nishina, Y., Davydov Splitting of Degenerate Lattice Modes in the Layer Compound GaS. *Phys. Rev. B* **1979,** *19*, 1312-1315.
34. Ni, Y. Y. W. Z. H.; Shena, Z. X.; Wang, H. M.; Wu, Y. H., Interference Enhancement of Raman Signal of Graphene. *Appl. Phys. Lett.* **2008,** *92*, 043121/1-043121/3.

35. Mak, K. F.; Lee, C.; Hone, J.; Shan, J.; Heinz, T. F., Atomically Thin $MoS_2$: A New Direct-Gap Semiconductor. *Submitted* **2010**.
36. Bagnall, A. G.; Liang, W. Y.; Marseglia, E. A.; Welber, B., Raman Studies of $MoS_2$ at High Pressure. *Physica* **1980,** *99B*, 343-346.
37. Sandoval, S. J.; Yang, D.; Frindt, R. F.; Irwin, J. C., Raman Study and Lattice Dynamics of Single Molecular Layers of $MoS_2$. *Phys. Rev. B* **1991,** *44*, 3955-3962.

**Figures and captions**

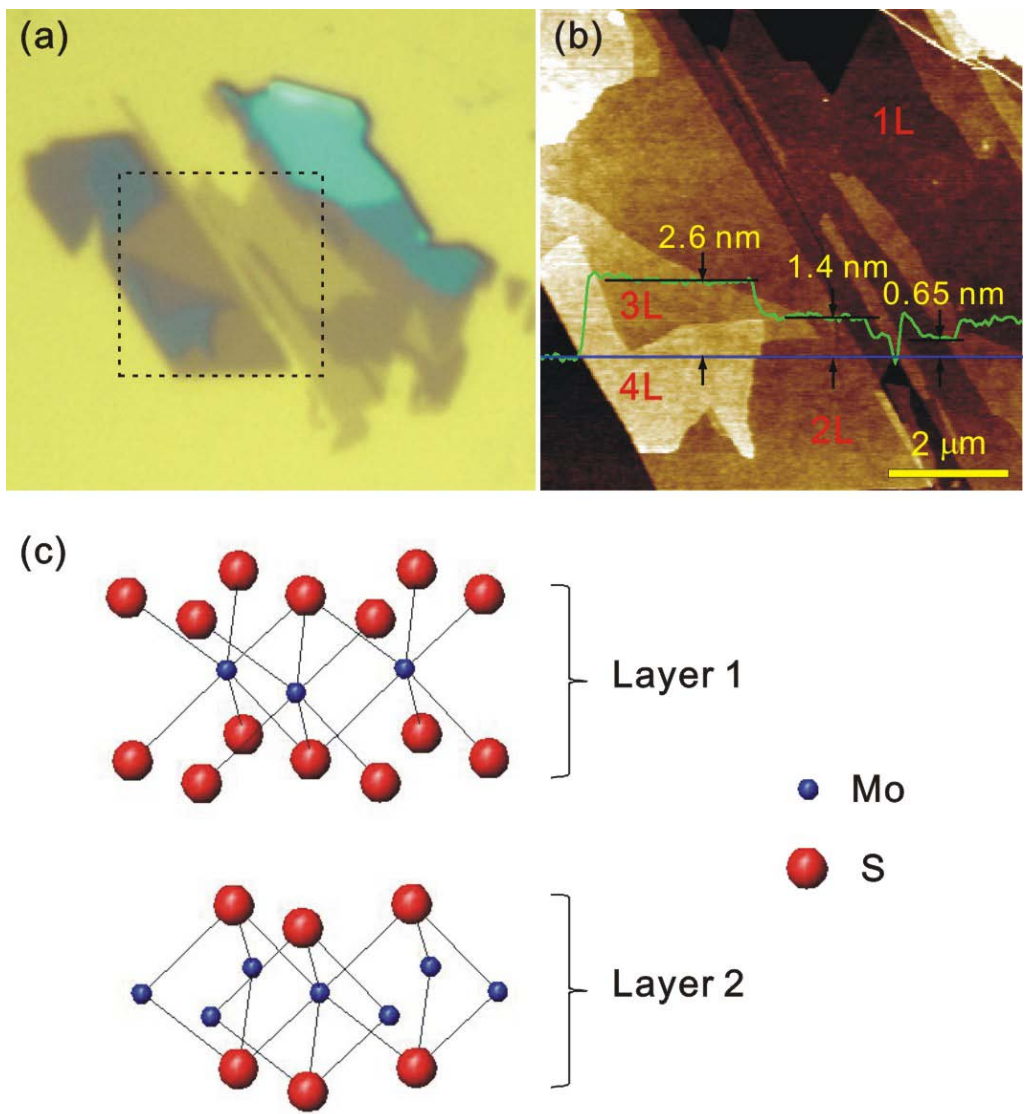

FIG. 1. (Color online) (a) Optical micrograph of thin $MoS_2$ films deposited on the $SiO_2$/Si substrate. (b) The AFM height image taken for the 8 × 8 μm$^2$ area indicated by dotted lines in (a). The thickness of each layer is shown by the height profile (in green) taken along the blue line in the AFM image. (c) A schematic model of the $MoS_2$ bilayer structure.

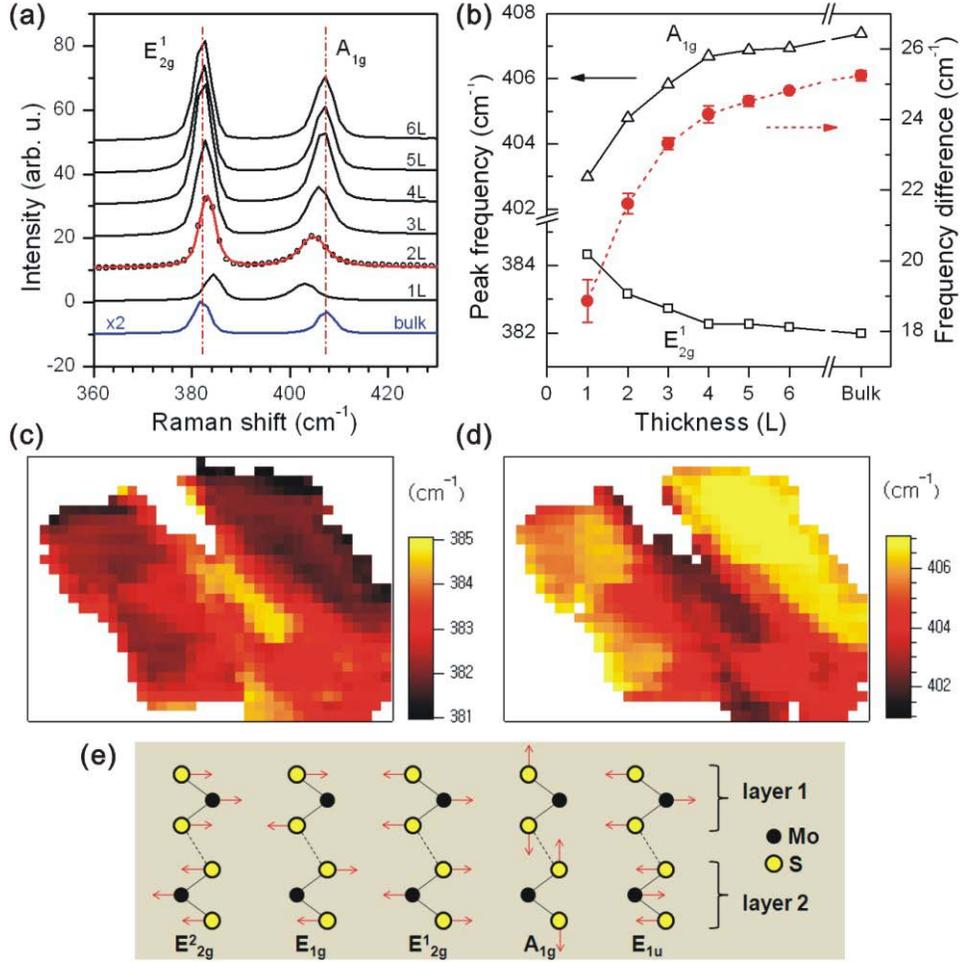

FIG. 2. (Color online) (a) Raman spectra of thin (nL) and bulk MoS$_2$ films. The solid line for the 2L spectrum is a double Voigt fit through data (circles for 2L, solid lines for the rest). (b) Frequencies of E$^1_{2g}$ and A$_{1g}$ Raman modes (left vertical axis) and their difference (right vertical axis) as a function of layer thickness. (c)&(d) Spatial maps (23 μm × 10 μm) of the Raman frequency of E$^1_{2g}$ (c) and A$_{1g}$ modes (d) for the sample in Fig. 1. (e) Atomic displacements of the four Raman active modes and one IR active mode (E$_{1u}$) in the unit cell of the bulk MoS$_2$ crystal as viewed along the [1000] direction.

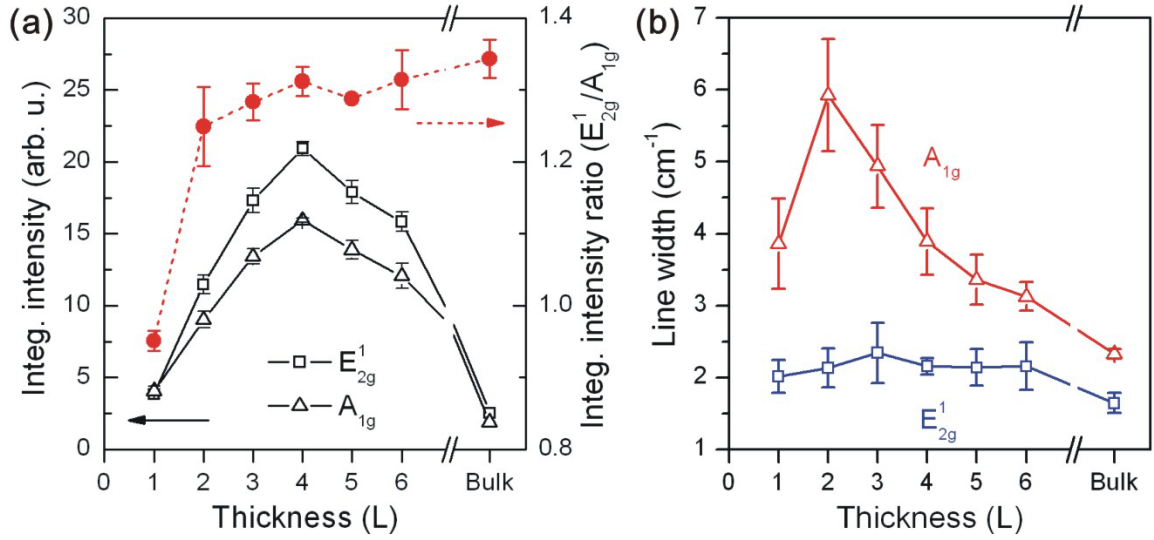

FIG. 3. (Color online) (a) Thickness dependence of integrated intensity (left vertical axis) and ratio of integrated intensity (right vertical axis) for the two Raman modes. (b) Linewidths Γ of the two modes as a function of sample thickness.

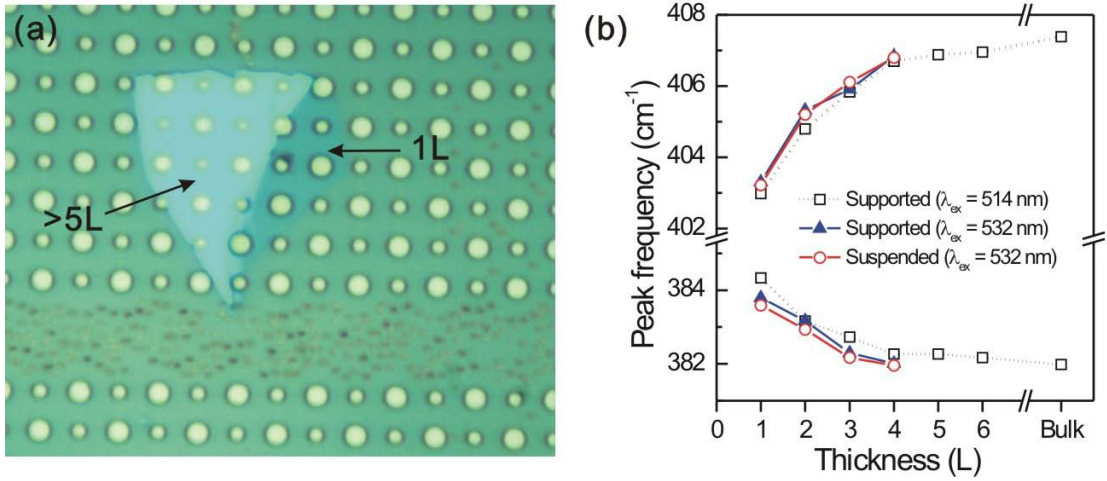

FIG. 4. (Color online) (a) Optical micrograph of single (1L) and few-layer (>5L) MoS$_2$ films placed on a patterned substrate with 1-and1.5-μm diameter holes. Raman measurements were carried out on both the supported and suspended portions of such samples. (b) Frequencies of E$^1_{2g}$ and A$_{1g}$ Raman modes as a function of layer thickness for MoS$_2$ samples supported on the substrate (blue triangles) and suspended over the holes (red circles). For comparison, the corresponding data for MoS$_2$ samples prepared on unpatterned substrates (from Fig. 2b, measured at a slightly different wavelength) are shown as open black squares.

**TOC Figure**

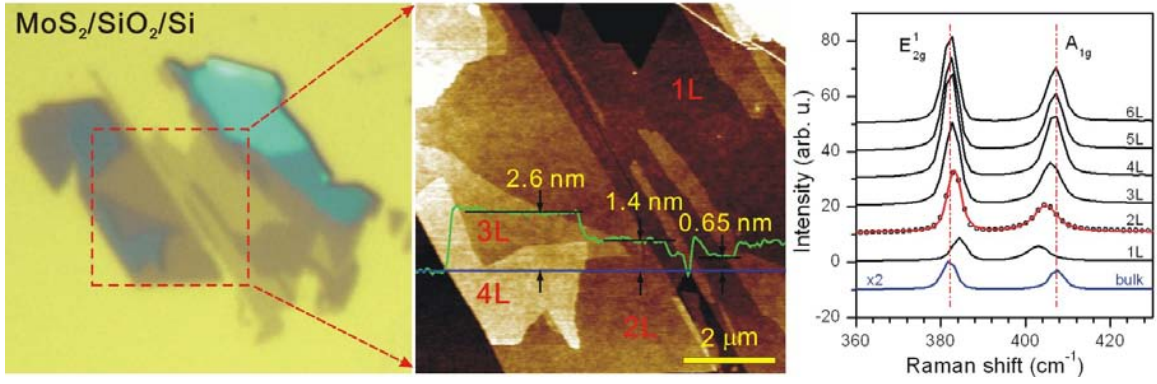